\documentstyle[prl,twocolumn,aps,graphics,epsf,rotate]{revtex}   
\begin{document}

\twocolumn[\hsize\textwidth\columnwidth\hsize\csname
@twocolumnfalse\endcsname

\title{Field Effect Magnetization Reversal in 
Ferromagnetic Semiconductor Quantum Wells}

\draft

\author{ Byounghak Lee$^1$, T. Jungwirth$^{2,3}$, A.H. MacDonald$^3$}
\address{$^1$Department of Physics, Indiana University,
Swain Hall W.\ 117, Bloomington IN 47405}
\address{$^2$Institute of Physics ASCR,
Cukrovarnick\'a 10, 162 53 Praha 6, Czech Republic}
\address{$^3$Department of Physics, The University of Texas at Austin, 
	Austin, TX 78712}
\date{\today}
\maketitle

\begin{abstract}
We predict that a novel bias-voltage assisted magnetization reversal process
will occur in Mn doped II-VI semiconductor quantum wells or heterojunctions 
with carrier induced ferromagnetism.
The effect is due to strong exchange-coupling induced subband mixing 
that leads to electrically tunable hysteresis loops.
Our model calculations are based on the mean-field theory of carrier induced 
ferromagnetism in Mn-doped  quantum wells and on a semi-phenomenological
description of the host II-VI semiconductor valence bands. 
\end{abstract}

\vskip2pc] 

\narrowtext

%%%%%%%%%%%%%%%%%%%%%%%%%%%%%%%%%%%%%%%%%%%%%%%%%%%%%%%%%%%%%%%%%%%%%
The strong dependence of transport properties in metallic multilayers 
on magnetic configuration has inspired a new field of research\cite{ballnature00},
spintronics, in which  the electron spin degree of freedom is exploited. 
Semiconductors that become ferromagnetic when doped with magnetic elements,
for example GaAs doped with Mn, are of special interest in this field
in part because of their compatibility with conventional semiconductor technology and 
because of the larger range of possibilities for the control of
device functionality.  Ferromagnetism has already been observed in such 
systems at relatively high temperatures ($T_c > $100K) \cite{ohnosci98} and 
strategies for obtaining $T_c$'s larger than 
room temperature are of a great current interest in this scientific 
community\cite{dietlsci00,matsumotosci01}.

Our interest here is in the magnetic properties of modulation doped 
quantum wells containing Mn local moments.
We expect\cite{leeprb00,breyprl00} these systems to be unusual ferromagnets because of 
the reduced dimensionality of the itinerant hole system and 
the possibility of using confinement effects and doping 
profiles to manipulate their magnetic properties.
For example, the ferromagnetic critical temperature can be tuned by
an external electric field \cite{leeprb00}, as demonstrated in recent 
studies of (In,Mn)As field effect transistors\cite{ohnonature99}.
In this Letter we study their hysteresis properties which will be, we predict, 
extremely sensitive to external bias voltages.
The interplay between local moment - itinerant hole exchange coupling, 
Zeeman coupling of the localized moments to external magnetic field, and 
the coupling of an external electric field to the orbital degrees of freedom 
of itinerant holes, makes it possible to vary the coercive field 
by over an order of magnitude with rather modest external electric fields.
As a result, we predict that the magnetization orientation in quantum
wells can be manipulated electrically without changing the 
magnetic field. 
Our calculations also suggest that capacitance measurements can be used to 
probe the magnetic state of biased ferromagnetic semiconductor quantum wells. 

The system we consider is a (Cd,Mn)Te quantum well grown in the 
$\langle 001\rangle$ crystal direction.
Unlike III-V host materials, where Mn acts as an acceptor,
the Mn concentration and the itinerant hole density can be varied 
independently in II-VI hosts.  The quantum effects discussed here are more
pronounced when only one or two 
subbands are occupied, a condition that can easily be achieved with sizeable
Mn concentrations in II-VI diluted magnetic semiconductor (DMS) quantum wells. 

The Hamiltonian of a DMS system is written using an envelope function 
description of  valence band electrons and a spin representation of
their kinetic-exchange interaction \cite{dmsreview} with d-electrons
\cite{bhattssc00} on the Mn$^{2+}$ ions:
\begin{equation}
{\cal H} = {\cal H}_m + {\cal H}_b + J_{pd} \sum_{i,I} {\vec S_I} \cdot {\vec
s}_i \; \delta({\vec r}_i - {\vec R}_I),
\label{coupling}
\end{equation}
where $i$ labels a valence band hole and $I$ labels a magnetic ion. In
Eq.~(\ref{coupling}), ${\cal H}_m$ describes the coupling of magnetic ions with
total spin quantum number $S=5/2$ to an external field,
$\vec S_I$ is a localized spin, $\vec s_i$ is a hole spin, and ${\cal H}_b$ is
a four-band envelope-function Hamiltonian \cite{luttingerpr55} for
the valence bands. 
The four-band Kohn-Luttinger model describes only the total
angular momentum $j=3/2$ bands, and is adequate at low hole densities
when spin-orbit coupling is large. 
The degeneracy between heavy-hole ($|j_z|=3/2$) and light-hole
($|j_z|=1/2$) bands at the $\Gamma$-point in the bulk is lifted by size
quantization effects in a quasi-two-dimensional system. 
The resulting heavy-light gap is larger than the Fermi energy, in the relevant range of hole 
densities, allowing only the two heavy-hole bands to be occupied. 
The heavy holes have their spin aligned along $\hat{z}$-axis 
($\langle 001\rangle$ crystal direction) so that the band electron spin 
matrix elements get smaller when the magnetization tilts away from
the growth direction.   This leads, as discussed below, to 
very strong magnetic anisotropy with easy axes along and opposite to 
the growth direction.  This anisotropy is reduced in magnitude but is
still large when mixing between 
heavy- and light-hole bands is fully accounted for, as we show below. 
The exchange interaction between valence band holes and localized moments 
is believed to be antiferromagnetic \cite{dmsreview}, i.e. $J_{pd}>0$.
For (Cd,Mn)Te, $J_{pd}\approx 0.06$~eV~nm$^{3}$. 

Our mean-field theory is derived in the spin-density-functional framework
and leads to a set of physically transparent coupled equations
\cite{jungwirthprb99}. The effective magnetic field seen by localized
magnetic ions is the sum of the external magnetic field and the kinetic
exchange coupling mean-field contribution from spin-polarized carriers:
\begin{equation}
{\vec H}_{eff}({\vec R}_I) = {\vec H}_{ext}
        - J_{pd} \langle {\vec s}({\vec R}_I) \rangle /  g \mu_B \; ,
\label{Heff}
\end{equation}
where ${\vec H}_{ext}$ is the external magnetic field,
$\langle {\vec s}({\vec R}_I) \rangle$ is the carrier spin density at Mn sites,
and $g=2$ is the g-factor of the local moments.
The mean polarization of a  magnetic ion is given by \cite{aharoni}
\begin{equation}
\langle {\vec S} \rangle({\vec R}_I) =
        S B_{S}\big(S H_{eff}({\vec R}_I) / k_B T\big)
        {\hat H_{eff}}({\vec R}_I)\; ,
\label{SI}
\end{equation}
where $B_S(x)$ is the Brillouin function and ${\hat H_{eff}}({\vec R}_I)$
is the unit vector along the direction of the effective magnetic field 
defined in Eq.(\ref{Heff}).
The itinerant-hole spin-density is determined by solving the 
Schr\"{o}dinger equation for holes which experience chemical quantum
confinement and electrostatic external potentials, 
a local-spin-density-approximation exchange-correlation potential,
and a spin-dependent kinetic-exchange potential $h(z)$.
The field $h(z)$ is non-zero only in the ferromagnetic state.
We assume that the magnetic ions are randomly distributed
and dense on a scale set by the free carrier Fermi wavevector
and that any desired density profile, $N_{Mn}(z)$, can be achieved during 
growth.
This allows us to take a continuum limit where
\begin{equation}
h(z) = J_{pd} \  N_{Mn}(z)\ \langle S \rangle(z).
\label{h}
\end{equation}
Results presented in the following paragraphs are based on a self-consistent
solution of Eqs.~(\ref{Heff})-(\ref{h}) in the zero temperature limit.
This approach does not account for disorder associated with randomness in the 
Mn spin locations, which could have some importance in general, 
or for direct interactions between Mn spins on neighboring lattice sites\cite{localcorrelation}.
There is however a large body of literature, mostly concerning the magnetic
field dependence of optical properties of paramagnetic DMS systems, that 
supports the reliability of these approximations for qualitative predictions. 

\begin{figure}[tb]
\epsfxsize=3.5in
\centerline{\epsffile{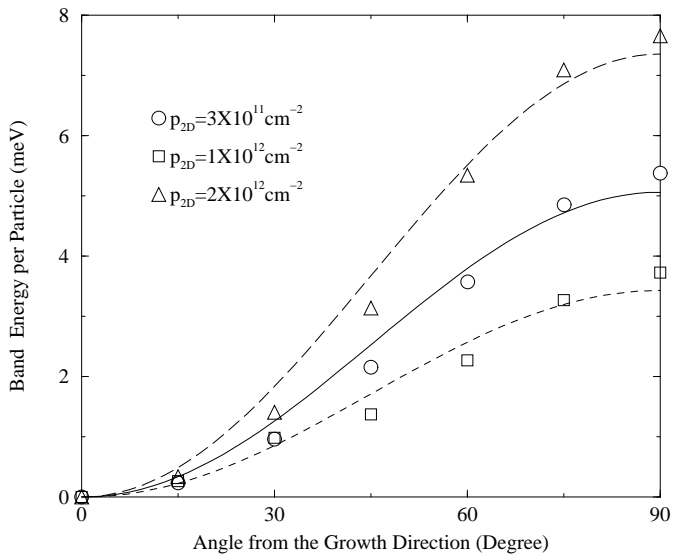}}
\vspace*{0.0cm}
\caption{Band anisotropy energy calculated from the 
self-consistent Hartree
approximation using the four-band Luttinger model
as a function of magnetization direction from the growth direction.
The anisotropy energy is defined as the energy difference compared to the 
growth direction orientation.
At the lowest density only one subband is occupied while two subbands are 
occupied at the two higher densities.
The curves fit the numerical data in a uniaxial form.}
\label{fig:aniso}
\end{figure}

The magnetization reversal properties of these quantum well ferromagnets
are quite unusual as we now establish.  We first discuss the 
dependence of the ground state energy on uniform magnetization orientation.
As explained above we expect that quantum confinement and valence band 
spin-orbit coupling should conspire to yield a large anisotropy energy.
In Fig.~\ref{fig:aniso} we plot the change in the total band 
energy per particle $E_{b}(\theta)$ as a
function of angle $\theta$ of the magnetization measured from the
growth direction for several 2D densities $p_{2D}$.
We assumed a (Cd,Mn)Te/(Cd,Zn)Te quantum well of the width 
$w = 10$~nm with the valence
band offset 150~meV and uniformly distributed Mn ions of density 
$N_{Mn}=6\times10^{20}$~cm$^{-3}$.   
The energy increase at non-zero angles, i.e.,
the growth direction easy-axis anisotropy, is consistent with experiment
\cite{kossackiphyse00}, and with the expectations based on 
$\langle 001\rangle$ growth 
quantum confinement effects explained above. 
If there were no heavy-light subband mixing at any orientation, the 
magnetic condensation energy would vanish for $\theta = \pi/2$ 
(the operators $s_x$ and $s_y$ vanish in the heavy-hole subspace)
and the anisotropy energy per electron would equal $J S N_{Mn}/2
\sim 45$~meV, much larger than the values obtained in Fig.~\ref{fig:aniso}.
This limit is reached, however, only when quantum well subband splittings
are much larger than $J S N_{Mn}$ and we are not close to this limit.
Nevertheless, the magnitude of the anisotropy energy we obtain is 
still several times larger than in bulk ferromagnetic
semiconductors\cite{abolfathprb01}.  
It is interesting to note that the anisotropy energy per electron in 
these ferromagnets is nearly three orders of magnitude larger than 
in cubic transition metal ferromagnets\cite{skomski}. 

\begin{figure}[tb]
\epsfxsize=3.5in
\centerline{\epsffile{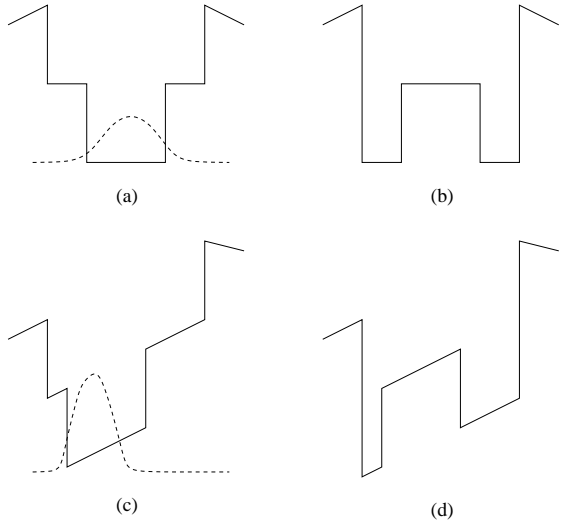}}
\vspace*{-2cm}
\caption{Schematic diagram of potentials in a DMS QW with external magnetic 
field, $0<|H_{ext}|<|H_r|$, which is oriented in opposition to the local spin 
moment.
$H_r$, the reversal field, is close to the field at which the lowest minority 
spin level crosses the majority spin energy level.
(a) and (b) show majority and minority spin potentials
without a bias potential, while  
(c) and (d) show majority and minority spin potentials with a bias potential 
applied.
The broken lines indicate the envelope wavefunctions.}
\label{fig:potential}
\end{figure}

For magnetization along the easy axis, heavy-light mixing is 
very weak permitting the use a single spin-split band model 
with effective mass $m_{\parallel}=0.25m_0$ and
out-of-plane mass $m_z=0.55m_0$. 
In this (Cd,Mn)Te/(Cd,Zn)Te quantum well, the spin-splitting is about 
three times larger than the splitting between the two lowest 
majority-spin subbands.
This enormous exchange gap between majority- and minority-spin states
is the origin of the peculiar magnetic reversal properties in biased 
DMS quantum wells that we now discuss.  Consider the effect of a field in the 
positive $\hat{z}$-direction on the state where all spins 
are polarized in the negative $\hat{z}$-direction.
Because of the antiferromagnetic $p-d$ coupling, spin-up holes are 
the majority carriers at zero external magnetic field. 
At the 2D hole densities of interest here, only the lowest spin-up subband is 
occupied and all spin-down subbands are empty, i.e., the itinerant system 
is fully polarized.
When the magnetic field is applied, direct Zeeman coupling to a local
moment competes with the local mean-field 
kinetic-exchange coupling which is proportional to the itinerant-hole 
spin-density. 
Since the carrier density is smaller at the edges of the quantum well
than at the center, spin reversal starts from the well edges.
This,  in turn,  creates an exchange barrier for the majority-spins
which effectively narrows the quantum well in which they reside.
At the same time an effective 
double-well potential develops for the minority spins as illustrated in 
Fig.~\ref{fig:potential}.
Note that because of the strong kinetic-exchange potential, 
the symmetric and antisymmetric minority-spin
states in the unbiased double-well remain nearly degenerate throughout the
whole meta-stable part of the hysteretic magnetization curve.
\vspace{-2cm}

As the external magnetic field increases,
the minority-spin energy levels are lowered and
the majority-spin energy levels are raised.
When the lowest down-spin energy level reaches the Fermi energy $E_F$,
holes start to occupy the down-spin states.  Our 
self-consistent calculations show that once this occurs,
the magnetization reversal is rapidly completed and only the 
uniform down-spin state is stable.

\begin{figure}[tb]
\epsfxsize=3.5in
\centerline{\epsffile{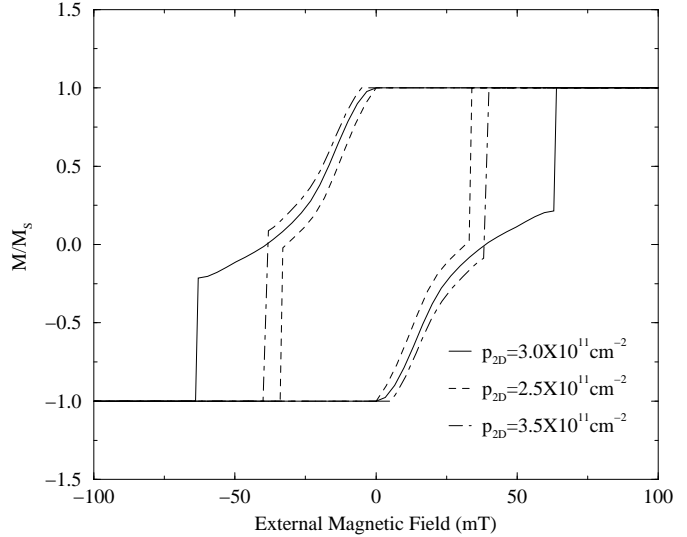}}
\vspace*{0.0cm}
\caption{Local moment hysteresis loops in a quasi-2D DMS system with 
different carrier densities.  
These curves were evaluated in the Hartree approximation.}
\label{fig:mvsb}
\end{figure}

This unusual process in which magnetization reversal and quantum
confinement effects are linked suggests that the hysteresis 
curve can be modified by applying an external electric field, as shown in 
Fig.~\ref{fig:potential}.  We can parameterize the electric
field between the front-gate and the quantum well, fixing the field
at the back side of the well, by the total two-dimensional 
carrier density.
We have calculated local moment magnetization loops for 
several carrier densities where $p_{2D}=3\times10^{11}$~cm$^{-2}$
corresponds to the balanced quantum well.
The results are plotted in Fig.~\ref{fig:mvsb}.
Two effects contribute to the change of the hysteresis loop with applied
electric field. First, the hole-density profile is compressed and moves
toward one edge of the well, permitting an abrupt localized
moment reversal throughout the depleted region of the quantum well. Second,
the increase (decrease) of the 2D hole density enhances (suppresses) the
kinetic-exchange coupling. 

The hysteresis loops shown in Fig.~\ref{fig:mvsb} imply that, at a
fixed magnetic field, the magnetization can be changed by applying an 
electric field.  
This possibility of narrowing or broadening hysteresis loops 
electrically is an attractive feature of ferromagnetic semiconductor
quantum well systems.  It is somewhat reminiscent of recent 
interest in current driven magnetization 
reversal\cite{curswitch} in metallic ferromagnets and 
could be extremely useful if other obstacles
(most obviously the low ferromagnetic 
transition temperatures) to the use of these systems in non-volatile memory
devices can be overcome.

\begin{figure}[f]
\epsfxsize=3.5in
\centerline{\epsffile{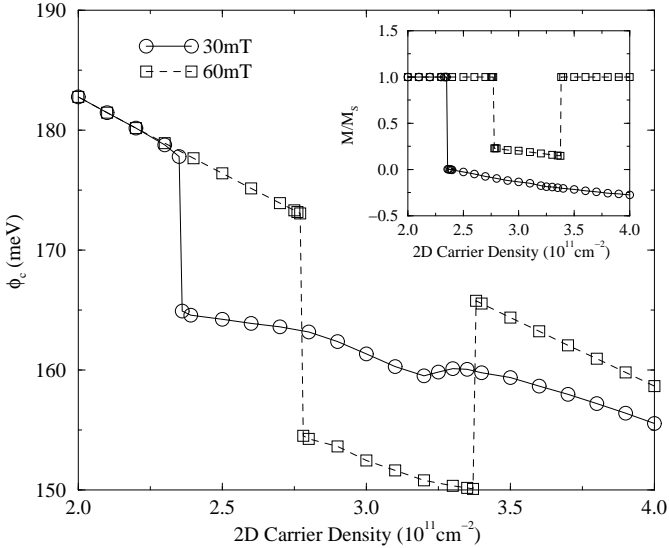}}
\vspace*{0.0cm}
\caption{Difference between the electric potential at a reference point 
outside the quantum well and the chemical potential of the two-dimensional 
electron system as a function of carrier density.  
inset: local moment magnetization as a function of carrier density.
These curves are calculated by starting in the metastable state of a balanced 
quantum well and varying the quantum well hole density.}
\label{fig:capacitance}
\end{figure}

The interplay between quantum confinement and magnetization reversal
in ferromagnetic semiconductor quantum wells is reflected in unusual 
magnetocapacitance effects.  The inverse capacitance can be written as
\begin{equation}
C^{-1} =  C_b^{-1} + C_c^{-1} \;,
\end{equation}
where $C_b^{-1}$ is the inverse geometric capacitance, and
$C_c^{-1}$ is the inverse 2D-channel capacitance.
The geometric capacitance depends only on the distance from the 
quantum well to the gate and on the barrier dielectric constant $\epsilon$,
and does not reflect the electronic structure of the 2D hole gas.
The inverse channel capacitance 
\begin{equation}
C_c^{-1}={d\phi_c \over dp_{2D}} =\frac{d}{dp_{2D}}\left(\frac{\overline{z}}{\epsilon}p_{2D}+
\frac{E_F}{e^2}\right)
\end{equation}
is the sum of the electrostatic term related to the density dependent
center of mass of the quasi-2D hole system and a thermodynamic
term originating from the concentration dependent Fermi energy.
This separation is physically sensible and convenient and 
can be made precise by the arbitrary choice of a reference position outside 
the quantum well; only the total capacitance is measurable.
Fig.~\ref{fig:capacitance} demonstrates the similarity of electric and magnetic 
responses to the electric bias field, indicating the magnetic state of this 
quasi-2D system can be detected electrically.  
These curves are calculated by starting in the 
partially magnetized state 
at the balanced density, $p_{2D} = 3 \times 10^{11} {\rm cm}^{-2}$; 
as explained previously the metastable state
persists to the largest reversal fields for balanced quantum 
wells.
When the density is varied, metastability is lost and the 
magnetization reversal is completed.  The discontinuity in magnetization is 
accompanied by a discontinuity in the electrochemical potential difference 
between channel and gate that  appears as a singularity in the capacitance.

In closing we remark that this unusual magnetization reversal process can 
be operative only if the system is always stable against coherent 
magnetization reorientation.  
We do not expect that this stability will hold for wider quantum wells, 
but believe that it does for the geometry considered here, as we now argue.  
Accounting for the coupling to an external field along
the easy axis to the Mn spins,
the total energy per electron as a function of the 
coherent spin orientation $\theta$ is:
\begin{equation}
E(\theta) = E_{b}(\theta)- \cos(\theta) g \mu_B S H_{ext} N_{Mn} \zeta(H_{ext}) w/p_{2D} \;,
\label{eq:coherentrotation}
\end{equation}
where 
$\zeta(H_{ext})=(N_{Mn}^{\uparrow}-N_{Mn}^{\downarrow})/
(N_{Mn}^{\uparrow}+N_{Mn}^{\downarrow})$ 
is the spatially averaged local moment polarization.
Our band anisotropy energy results, summarized in Fig.~\ref{fig:aniso}, 
are accurately described by a uniaxial form, $E_{b}(\theta) = K \sin^2(\theta)$ with 
$K = 5 {\rm meV}$ per electron for $p_{2D} = 3 \times 10^{11} {\rm cm}^{-2}$.
If $K$ were constant and the $\zeta(H_{ext})$ were equal to one, the 
coherent rotation field would be $H_{c} = 2 K/ (g \mu_B S N_{Mn} w/p_{2D}) \sim 17 {mT}$.  
However $\zeta$ is reduced in the spatially inhomogeneous state and 
$K$ will be increased by the effectively narrower quantum well in which 
the majority spin electrons are confined.  
These two effects will combine to make the unusual 
reversal process discussed in this paper an operative one for quantum wells 
narrower than $\sim10 {\rm nm}$.

We acknowledge useful discussions with Tomasz Dietl, Jacek K. Furdyna,
J\"{u}rgen K\"{o}nig, Hideo Ohno, and John Schliemann. The work was supported
by DARPA, Indiana 21st Century Fund, Welch Foundation, and by the Ministry of 
Education of the Czech Republic under grant OC P5.10.

\end{document}